%% file: main.tex
\documentclass[runningheads]{llncs}
\usepackage{graphicx}

\usepackage{amsmath}
\usepackage{hyperref}       
\usepackage{url}            
\usepackage{booktabs}       
\usepackage{amsfonts}       
\usepackage{nicefrac}       
\usepackage{microtype}      
\usepackage{graphicx}
\usepackage{subcaption}
\usepackage{multirow}
\usepackage{numprint}
\usepackage{threeparttable}
\usepackage{etoolbox}
\usepackage{pgfplots}

\usetikzlibrary{calc}
\usetikzlibrary{positioning}
\usetikzlibrary{decorations.pathreplacing}

\usepackage{algorithm}
\usepackage{algcompatible}
\usepackage{float}
\usepackage{csquotes}

\usepackage{expl3}

\hypersetup{breaklinks=true}
\usepackage{cite}

\algblockdefx{FORALLP}{ENDFAP}[1]%
  {\textbf{for all }#1 \textbf{do in parallel}}%
  {\textbf{end for}}

\appto\TPTnoteSettings{\footnotesize}
\pgfplotsset{compat=1.8}

\npdecimalsign{.}
\nprounddigits{3}
\npfourdigitnosep

\begin{document}

\title{Shareable Representations for Search Query Understanding\thanks{Supported by Amazon.}}
%
%
\author{Mukul Kumar\inst{1} \and
Youna Hu\inst{1} \and
Will Headden\inst{1} \and
Rahul Goutam\inst{1} \and
Heran Lin\inst{1} \and
Bing Yin\inst{1}}


%
\authorrunning{M. Kumar et al.}
%
\institute{Amazon Search Science and AI}
\maketitle              
\begin{abstract}

\input{00_abstract.tex}

\keywords{Transfer Learning \and Phrase Embeddings  \and Short Text Classification \and Scaling Deep Learning}
\end{abstract}

\input{01_introduction.tex}

\input{02_related_work.tex}

\input{04_data_baselines.tex}

\input{03_model_training.tex}

\input{05_experiments_results.tex}

\input{06_production.tex}

\input{07_conclusion.tex}

\bibliographystyle{splncs04}
\bibliography{references}

\end{document}

%% file: 00_abstract.tex

Understanding search queries is critical for shopping search engines to deliver a satisfying customer experience. Popular shopping search engines receive billions of unique queries yearly, each of which can depict any of hundreds of user preferences or intents. In order to get the right results to customers it must be known queries like ``inexpensive prom dresses'' are intended to not only surface results of a certain product type but also products with a low price. Referred to as query intents, examples also include preferences for author, brand, age group, or simply a need for customer service. Recent works such as BERT have demonstrated the success of a large transformer encoder architecture with language model pre-training on a variety of NLP tasks. We adapt such an architecture to learn intents for search queries and describe methods to account for the noisiness and sparseness of search query data. We also describe cost effective ways of hosting transformer encoder models in context with low latency requirements. With the right domain-specific training we can build a shareable deep learning model whose internal representation can be reused for a variety of query understanding tasks including query intent identification. Model sharing allows for fewer large models needed to be served at inference time and provides a platform to quickly build and roll out new search query classifiers.

%% file: 01_introduction.tex
\section{Introduction}
\label{sec:intro}


Query intent information can be used to improve features throughout the search engine including filtering and ranking of search results. This information includes knowing that with a query like \emph{cheap iphone cases} a customer is searching for a product type of \emph{iphone case} and also wants results that are ``cheap''. We refer to such indicators as \emph{query intents}.  In this work we focus on three intents: help, adult and low average selling price. A query with help intent implies a need for customer service rather than trying to buy something. Queries with help intent include \emph{help with fire tv repair} and \emph{how to install turbo tax on mac pro}. Adult intent implies just as much, that a customer is interested in products that are adult in nature. A query with a low average selling price intent indicates a desire for products in a certain price range.

Although we have identified three query intents here, there are hundreds more and the number is constantly increasing as our product offerings diversify and our customer needs expand. 
We investigated how we could use the latest deep learning techniques to build a large number of query intent classifiers in a resource efficient manner. Specifically, we explored how to leverage the benefits of larger models (100M+ parameters) in a model serving context where low latency is required in a way that is scalable and cost-effective. Specifically, we investigated the use of a transformer encoder architecture because it was found to produce state-of-the-art performance for many NLP tasks \cite{devlin2018bert}. We evaluated models that were pre-trained on general tasks like language modeling as well as on a task specific to the search domain: product category classification for queries. We will describe how shareable models can be used to build a service where partner teams can provide their own data and quickly build classifiers to suit their needs. The query intent tasks we chose to evaluate can be considered example use cases that a partner might approach such a service with. Our contributions are as follows:
\begin{enumerate}
\item We demonstrate that a transformer encoder architecture can be fine-tuned for search query classification tasks in a way that outperforms classic approaches.
\item We demonstrate that if we pre-train a large model on a domain-specific task (e.g. product category classification for queries) we can reuse layers from the model to produce a representation for search queries that can be used for other classification tasks without having to separately fine-tune every parameter in the model on each task.
\item By separating a large deep learning model into components that are shareable and those that are task-specific, we can enjoy the improved prediction accuracy that comes from a large model trained on an extensive dataset without incurring the additional infrastructure costs of hosting a separate large model for each of potentially hundreds of tasks.
\end{enumerate}

%% file: 02_related_work.tex
\section{Background and Related Work}
\label{sec:related}

We first describe the latest efforts to create an embedding (or vector representation) for text that can be used in a variety of downstream tasks. Then we cover prior work in using an extreme classification task to train a model to differentiate between classes not seen during training. This technique is referred to as zero-shot learning. Finally, we discuss product category classification of queries, which is the extreme classification task we chose to use as our domain-specific pre-training task.

\subsection{Representation Learning for Text}

Using embeddings to represent words \cite{NIPS2013_5021,Pennington14glove,DBLP:conf/emnlp/ChenC18} and phrases \cite{DBLP:journals/corr/abs-1806-06259,Kiros:2015:SV:2969442.2969607} has proven useful for natural language understanding. Creating and using embeddings with standard deep learning architectures such as CNNs has also been researched \cite{queryintentcnn,D181093}. It was recently demonstrated through transfer learning \cite{cui2018transfer} that word-level and phrase-level representations can be learned via encoder-decoder models trained on a language modeling task \cite{45446,P18-1031}. The output of an encoder trained in this fashion has been proven to be an effective representation for text in many standard natural language understanding tasks. In ELMo, the encoder has a BiLSTM architecture that is trained on text in both directions \cite{peters2018deep}. Another architecture, the transformer, utilizes self-attention layers in an encoder-decoder architecture that has become popular for a variety tasks including machine translation and natural language processing \cite{attention2017}. The Universal Sentence Encoder \cite{cer2018universal} demonstrated that a transformer encoder could be trained to create representations for text that cluster sentences with similar meaning together across a variety of domains. The training for this encoder is not domain-specific. With BERT (Bidirectional Encoder Representations from Transformers), the encoder is also a transformer encoder architecture and is trained on NLP tasks \cite{devlin2018bert}. Two major innovations from this paper were a masked language modeling training task as well as dividing training into non-domain-specific ``pre-training'' tasks and domain-specific training tasks. We use a transformer encoder architecture similar to BERT as a starting point for building a representation that works across a variety of search related tasks.

\subsection{Zero-shot Learning and Extreme Classification}

Supervised learning has been shown to work well on classes that were seen during training and test time. Learning representations with a classification task that allows one to determine if two examples belong to a class not seen at training time is known as zero-shot learning. This approach has been demonstrated to work across a number of domains on a variety inputs including text, audio, and images \cite{NIPS2013_5027}. Inspired by this, we investigated to see whether the representation created by an extreme multi-classification problem would create representations that would make it easy to determine if a query belong to a class that was not originally seen at training time. Applying this to our specific task, if the encoder output could sufficiently differentiate queries, we could then build binary classifiers with the encoder output as the feature input for query intent identification. Using a classification task \cite{D18-1070} to create representations was also shown to be effective with Universal Sentence Encoder \cite{cer2018universal} as well as in the audio domain with Deep Speaker \cite{deepspeaker}. Deep Speaker attempted to create embeddings that could be used for speaker identification by using triplet loss to create an embedding space that properly clustered samples from the same speaker closer together than samples from different speakers. The model was first pre-trained on a classification task prior to fine-tuning with triplet loss. Our takeaways from this work was a classification objective could be used to create representations that would allow for samples with the same label to be clustered closer together when evaluating their vector representations using cosine similarity or other vector similarity metrics. We attempted to see if classification could provide this clustering property for the Search domain \cite{DBLP:journals/corr/ZamaniC17,product-query-classification} except in our case the search queries would have noisy and sparse labeling data. When considering a task to create an embedding space for search queries, we chose a task that classifies queries according to their product type. In our case, product type was represented by a hierarchical taxonomy of product categories. This seemed a reasonable approach because embeddings have been used to represent both queries and product titles in the search domain \cite{Grbovic:2018:RPU:3219819.3219885} as well as in a multi-task context \cite{D18-1484}.

\subsection{Training Strategies Involving Freezing Model Parameters}

Different layers of a deep learning model can learn different aspects of a particular domain and training strategies have been developed that selectively update only a subset model parameters when training for specific tasks. Pre-training a model and then only training the last layer or last few layers was shown to be effective in computer vision \cite{NIPS2014_5347}. ULMFit expanded on this by introducing gradual unfreezing of parameters layer by layer during training starting from the last layer and gradually moving back. This technique better held onto knowledge that was acquired in any pre-training tasks applied on the model\cite{P18-1031}. This indicated that for some multi-layered architectures early layers are more general and the last layers are more task-specific. The findings were similar regarding more recent encoders for word representations such as ELMo \cite{evalemb}. BERT was shown to have similar properties in that some layers could be shown to learn language features and some layers learned task-specific information\cite{evalbert}.


%% file: 04_data_baselines.tex
\section{Datasets}
\label{sec:datasets_baselines}


Search queries associated with each of the query intents we chose to investigate were collected from our logs. All user queries and data are either anonymized, aggregated, or otherwise not identifiable in order to preserve customer privacy. The following is a description of the data for each intent.

\subsection{Help Intent}

On the Amazon website, there is a separate portal for customers to ask for help and search for customer service related content. Nonetheless, customers can and do enter customer service queries into the main Amazon search bar and expect customer service content to surface when they do. To facilitate this we set out to train a classifier to identify likely help intent queries. We collected queries for both help intent and product intent based on user action that associated a query with the part of the website intended for help queries as well as regular non-help (product) searches. 
There were a total of 2,511,997 queries split between help and non-help. Because there were many fewer help samples than non-help samples, the help data was split into training, test, and validation sets and then down sampled so the non-help data matched the size of the help data set.


\subsection{Adult Intent}

By analyzing user behavior in our logs we were able to identify a set of queries that had an adult intent. If a query was associated with a certain number of interactions with a product that had an adult classification then we consider the query an ``adult'' query. Our dataset contains a total of 2,903,319 queries split evenly between adult and non adult classes that we use for experiments and evaluation.


\subsection{Low ASP Intent}

We use the historical price mean and standard deviation of the purchase price for each query to identify low average selling price queries (low ASP). For our experiments and evaluation, we took a subset of Amazon queries associated with fashion, jewelry, shoes, and luggage queries based on user action and divided the set into two categories: either \textless \$10 or \textgreater \$10. There were 3,006,440 queries in total. Since the fraction of \textless \$10 queries is smaller, we down-sampled the \textgreater \$10 queries to be the same as the number of \textless \$10. 


%% file: 03_model_training.tex
\section{Model Architecture \& Training Procedure}
\label{sec:model_training}

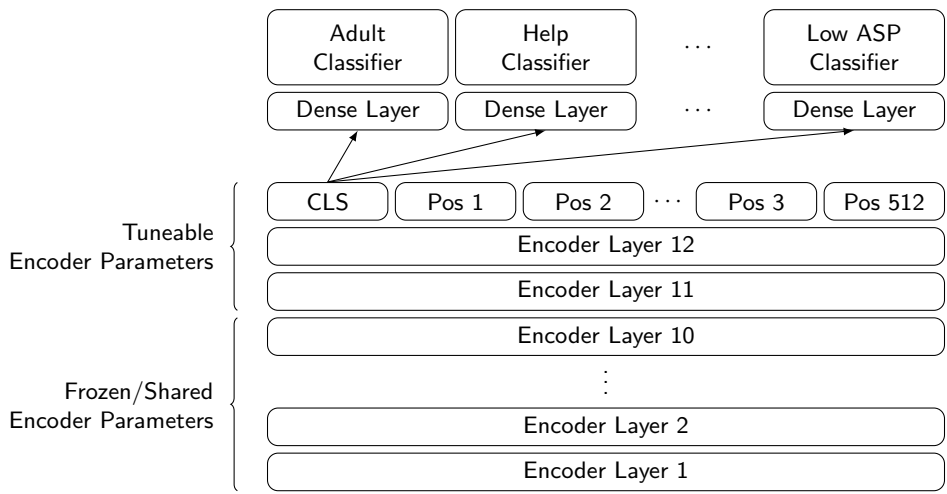
\begin{figure}[t]
\centering
{\sffamily
\begin{tikzpicture}[font=\footnotesize]

    \foreach \n/\y in {1/1,2/1.6,10/2.8,11/3.4,12/4} {
        \node[rectangle, rounded corners, draw, minimum width=9cm, minimum height=0.5cm] (encoder-\n) at (5.5, \y) {Encoder Layer \n};
    }
    \node (encoder-ellipsis) at (5.5, 2.3) {$\vdots$};

    \node[rectangle, rounded corners, draw, minimum width=1.6cm, minimum height=0.5cm] (cls) at (1.8, 4.6) {CLS};
    \foreach \n/\x in {1/3.5,2/5.2,3/7.5,512/9.2}
        \node[rectangle, rounded corners, draw, minimum width=1.6cm, minimum height=0.5cm] (pos-\n) at (\x, 4.6) {Pos \n};
    \node (embedding-ellipsis) at (6.35, 4.6) {$\cdots$};
    
    \foreach \n/\x in {1/2.2,2/4.7,3/8.8} {
        \node[rectangle, rounded corners, draw, minimum width=2.4cm, minimum height=0.5cm] (dense-\n) at (\x, 5.8) {Dense Layer};
    }
    \node (dense-ellipsis) at (6.75, 5.8) {$\cdots$};

    \foreach \name/\n/\x in {Adult\\Classifier/1/2.2,Help\\Classifier/2/4.7,Low ASP\\Classifier/3/8.8} {
        \node[rectangle, rounded corners, draw, minimum width=2.4cm, minimum height=1cm, align=center] (classifier-\n) at (\x, 6.65) {\name};
    }
    \node (classifier-ellipsis) at (6.75, 6.65) {$\cdots$};

    \foreach \n in {1,2,3}
        \draw[-latex] (cls.north) -- (dense-\n.south);
        
    \draw[decoration={brace,raise=0.4cm},decorate] (1, 0.75) -- node[left=0.6cm, align=right] {Frozen/Shared\\Encoder Parameters} (1, 3.05);
    \draw[decoration={brace,raise=0.4cm},decorate] (1, 3.15) -- node[left=0.6cm, align=right] {Tuneable\\Encoder Parameters} (1, 4.85);
\end{tikzpicture}
}
\caption{Transformer Encoder with 12 layers with frozen and tuneable layers.}
\label{fig:encoders}
\end{figure}

\subsection{Core Architecture}

We use a transformer encoder \cite{devlin2018bert,attention2017,mtdnn} as the starting point for a model that has common layers and also task specific layers. We also use the English lowercase WordPieces used in \cite{nmt2016} to provide the initial tokenization of input queries. The transformer encoder consists of 12 self-attention layers. Each encoder layer has 12 attention heads. The output of the model we use is a 768 dimension embedding corresponding to the CLS token described in the literature. Since we are focused on phrase-level representations, we ignored other subword-level embeddings produced by the architecture. Our goal was to train for the search domain generally so we can adapt to new tasks later. This is different from MT-DNN \cite{mtdnn} in that we allow for varying layers in transformer encoder itself to be fine-tuned for specific tasks after domain specific training was completed and the model deployed. We imagine the initial model being used in an ``as a service'' context where a common set of parameters are shared among multiple independent tasks that can be added and removed as needed by partner teams.

\algnewcommand\algorithmicforeach{\textbf{for each}}

\algblockdefx{FORALLP}{ENDFAP}[1]%
  {\textbf{ for all } \textbf{do in parallel}}%
  {\textbf{end for}}

\algblockdefx{FORALL}{ENDFA}[1]%
  {\textbf{for all}#1 \textbf{do}}%
  {\textbf{end}}

\begin{algorithm} 
\caption{Domain-specific Training and Task-specific Fine-tuning}
\label{alg:training}

\begin{algorithmic}[H]

\STATE{1. Intitialize encoder parameters $\theta$ with BERT parameters}
\STATE{2. Domain-specific training via product category classification}
\STATE{3. Freeze parameters $\theta$ for specified number of encoder layers}
\FORALL{$\space\; epoch \in Epochs $}
\FORALL{$\space\; batch \in Batches $}
    \STATE $emb \leftarrow encoder(batch)$
    \STATE $out \leftarrow task(emb)$
    \STATE $L(\theta) \leftarrow compute\_task\_loss$
    \STATE $compute\_gradient: gradient(\theta)$
    \STATE $update\_model: \theta = \theta - \epsilon(gradient(\theta))$
\ENDFA
\ENDFA
\end{algorithmic}

\end{algorithm}

\subsection{Domain-specific Training}
As shown in \autoref{alg:training}, we first initialized this model with BERT parameters \cite{devlin2018bert} and then trained this encoder to map search queries to a hierarchical taxonomy of product categories. The tree-like structure places more general product categories like ``Sports and Outdoors'' at the top of the tree with child categories being more specific. ``Athletic Clothing'', ``Exercise and Fitness'', and ``Team Sports'' are examples of child nodes of ``Sports and Outdoors''.

We map queries to product categories by examining user actions. For example, if in response to the query ``running shoes'' a user clicked on the product ``Nike Men's Revolution 4 Running Shoe'', we would associate the query with the product category for this product. In this case, the product category is ``Mens: Shoes: Athletic: Running: Road Running''. By learning on this task we were conditioning the BERT encoder model to short text input (search queries) and also training an embedding space that separated queries by product category. Variations of this have been described in other works as well \cite{DBLP:journals/corr/ZamaniC17}. 

We frame this problem as an extreme multi-label classification problem: a binary classification for each of the possible product type categories. We consider this ``extreme'' classification because there are tens of thousands of product categories. We train the model from historical user engagement; each product has its category assignments, and we can aggregate over items over which the customer engaged (clicked) for a given query. We do not explicitly add hierarchical constraints to enforce the score of parent product categories to be no less than the ones for child product categories, but in practice we found the hierarchical property is violated infrequently in the prediction given sufficient training data. We use binary cross-entropy (\autoref{eq:loss}) as the loss function, which is the standard choice for multi-label classification. 

\begin{equation}
\label{eq:loss}
L(\mathbf{y}, \hat{\mathbf{y}}) = - \sum_i y_i \log \hat{y}_i + (1 - y_i) \log (1 - \hat{y}_i)
\end{equation}

We use the number of user actions associated between a given query and a product as a filter for our training data. The higher number of user actions associated with a query and product category, the more likely the association is. It was found in other query to product category classification experiments that click signal can be noisy and sparse for some queries and product categories, especially for those that appear less frequently in the data set. We address this by training on a year's worth of United Kingdom marketplace data and only used query to product category pairs with a at least a certain number of user actions. Additionally, we take into account the type of user action. For example, we might consider a user action where a purchase occurred to be a stronger signal then a user action that was just a click. We also use a batch size of 4000 and train on a total of 4 billion samples. Although this is much larger than batch sizes recommended by \cite{devlin2018bert}, we found this resulted in a model that achieved better validation accuracy then smaller batches. We also used standard regularization methods (dropout and batch normalization).

\subsection{Task-specific Fine-tuning}

As stated above, we used a transformer encoder that was initialized with BERT parameters and then trained with a product category classification task for our experiments. We froze all the parameters of the encoder and trained a simple binary classifier with the encoder providing the input representation. The output of the encoder is a single dense layer. Similar to BERT, we extract the first 768 dimension embedding (called CLS in \autoref{fig:encoders}) of the dense layer and use this as a representation of the entire query. The binary classification layer was a single layer feed-forward network with an input size of 768 and an output size of 2 which corresponds to 1536 additional parameters in total. We investigated if fine-tuning just these parameters was sufficient to train a good binary classifier for our 3 query intents. We also investigated whether partially freezing the parameters in the encoder (e.g. the first 10 transformer layers instead of all 12) to investigate what kind of results a partial fine-tuning of the encoder would produce. As a baseline, we also trained the encoder in the way described in BERT \cite{devlin2018bert} where all parameters in the encoder were fine-tuned on each query intent task. In general we observed that model training would typically converge within an hour. The motivation for freezing various layers of the model came from works \cite{evalemb,evalbert} that demonstrated that different layers of a deep model for text representations can learn different information regarding the text. By freezing some of the information in the earlier layers and making information learned in the later layers tuneable to the task \cite{evalemb,evalbert,D18-1179}, we hoped we could find a good balance between general training and task-specific fine-tuning.

From an operations standpoint this is far more scalable then joint training on multiple tasks. If we exposed a model as a service and require joint task learning of the common model for all customer tasks we would run into two problems. First, we would have to retrain the entire initial model with each new task, second, we might introduce regressions in other tasks as we add training for new ones. By exposing different layers in a frozen model for input into downstream tasks, selection of input can be task specific and allows for the use of dedicated encoder layers when necessary.

%% file: 05_experiments_results.tex
\section{Shareable Embedding Experiments \& Results}
\label{sec:experiments_results}

\begin{table}[t]
\caption{Experiment Results on Validation Sets}
\label{tab:results-validation-set}
\centering
\begin{threeparttable}
\begin{tabular}{llllr N{1}{3} N{1}{3} N{1}{3}}
\toprule
Model & Pre-Train & Frozen L. & Train L. & Train Params    & \multicolumn{1}{c}{Help} & \multicolumn{1}{c}{Adult} & \multicolumn{1}{c}{Low ASP} \\
\midrule
DNN + $n$-grams & Random    & 0   & 1         & 11,167,234           & 0.90502         & 0.9051         & 0.7505          \\
Trans Enc (1)   & Random    & 0   & 1         & 30,926,594           & 0.92555         & 0.9280         & 0.7589          \\
Trans Enc (12)\tnote{a} & BERT      & 0   & 12        & 108,893,186          & 0.94015         & 0.9402         & 0.7714          \\
Trans Enc (12) & DST\tnote{b}       & 0   & 12        & 108,893,186          & \textbf{\numprint{0.9447}}& 0.9433         & 0.7834          \\
Trans Enc (12) & BERT      & 10  & 2         & 14,177,282           & 0.938         & 0.9328         & 0.7714          \\
Trans Enc (12) & DST       & 10  & 2         & 14,177,282           & 0.938         & \textbf{\numprint{0.9442}}& 0.7864          \\
Trans Enc (12) & BERT      & 11  & 1         & 7,089,410            & 0.9373         & 0.9246         & 0.7625          \\
Trans Enc (12) & DST       & 11  & 1         & 7,089,410            & 0.9385         & 0.9384         & \textbf{\numprint{0.7878}} \\
Trans Enc (12) & BERT      & 12  & 0         & 1,538                & 0.89408         & 0.8432         & 0.6748          \\
Trans Enc (12) & DST       & 12  & 0         & 1,538                & 0.92771         & 0.9239         & 0.7563          \\
\bottomrule
\end{tabular}
\begin{tablenotes}
    \item[a] Trans Enc (12): transformer encoder with 12 self-attention layers
    \item[b] DST: domain-specific training
\end{tablenotes}
\end{threeparttable}
\end{table}


\begin{table}[t]
\caption{Results on Test Set For Best Models W/ Some Frozen Parameters}
\label{tab:results-test-set}
\centering
\begin{tabular}{lllll N{1}{3} N{1}{3} N{1}{3}}
\toprule
Task & Model & Pre-Training & Frozen L. & Train L. & \multicolumn{1}{c}{Prec.} & \multicolumn{1}{c}{Recall} & \multicolumn{1}{c}{Acc.} \\
\midrule
Help & Trans Enc (12) & DST & 10 & 2 & 0.9568 & 0.9158 & 0.9360 \\
Adult & Trans Enc (12) & DST & 10 & 2 & 0.9522 & 0.9426 & 0.9494 \\
Low ASP & Trans Enc (12) & DST & 11 & 1 & 0.8203 & 0.7835 & 0.8059 \\
\bottomrule
\end{tabular}
\end{table}

\subsection{Setup}

Our investigation involved the evaluation of three architectures: a standard deep neural network (DNN), a single-layer encoder with a transformer architecture \cite{attention2017} that takes WordPiece as input\cite{nmt2016}, and a transformer encoder that was initialized via BERT pre-training that also has WordPiece input.

The DNN and the single layer transformer encoder served as our baselines. The single-layer DNN provided a good demonstration of product category prediction performance for a model with a relatively small number of parameters (less than 10M). It takes as input an embedding that represents the average of unigrams, bigrams, trigrams, and word-grams that can be generated from a query. Using $n$-grams for feature input has proven effective in many previous works \cite{bojanowski2017enriching,E172068}. The second baseline model we used was a single-layer transformer encoder with random initialization, unlike the DNN, this model used WordPiece for input featurization.

The architecture for our shared representation model is a transformer encoder with a total of 12 self-attention layers. We used two versions, a BERT pre-trained model and a BERT pre-trained model that was then trained for the search domain via product category classification. We experimented with fine-tuning only specific layers in the architecture. In order for parts of a model to be shared between various models, they need to be frozen and treated as essentially black box features that are input for downstream models. In order to provide this black box abstraction, we froze model parameters during task training in the following manner. Starting from the first layer in the model, we would freeze the first $N$ encoder layers and allow the final layers to be fine-tuned to specific tasks. The experiments also included the case where none of the layers in the transformer encoder were trainable as well as the case where all layers were trainable. In this way, we evaluated the performance of the transformer encoder based on how it was recommended to do so in the original work\cite{devlin2018bert} as well as in scenarios where the parameters are shared between in-domain tasks. For consistency, we used the following hyper-parameters in all experiments: batch size of 500, 1000 steps per epoch, and a learning rate of $3\times10^{-5}$. The transformer-based architectures all had identical input featurization with 12 WordPieces generated per query (e.g. all transformer architectures). We also used the same dropout and normalization methods for all transformer encoders. In summary, the overall process is to: 1) initialize with BERT pre-training parameters 2) carry out domain-specific training via product category classification 3) fine-tune on query intents on a small set of unique layers.


\subsection{Observations}

\paragraph{\textbf{Overall}} The performance for the transformer encoder with domain-specific training (DST) was generally better than the performance for the model that was initialized with BERT parameters. Full results are in \autoref{tab:results-validation-set}. The product category classification task looks to serve as an excellent domain adaptation task. Additionally, we saw differences in performance depending on how many parameters were frozen. We saw that depending on the task and the dataset, a different number of parameters would need to be frozen in order to achieve the best overall prediction accuracy. This is expected depending on how different the task-specific dataset is from the DST training data. We did find that freezing the first 10 layers and allowing variable parts of the rest of the model to be tuned to a specific task tend to have competitive results. This means the first 10 layers are a good candidate for a common model with frozen parameters and used as input for downstream task specific models.

\paragraph{\textbf{Training}} During model training, all models reached peak accuracy on the validation set within 10 epochs and began overfitting to the training data. We took the best validation set accuracy of the first 10 epochs as our evaluation metric. With the hyper-parameters specified above, the DNN took around 1 minute per epoch to train, the transformer encoder with 1 layer took 2 to 3 minutes per epoch, and the transformer encoders with 12 layers took roughly 10 minutes per epoch. \autoref{tab:results-validation-set} shows full results. We found freezing different parts of the transformer encoder did result in some differences in prediction accuracy, as show in \autoref{tab:results-test-set}. \autoref{tab:results-test-set} also reflects differences in prediction accuracy that comes from initializing the transformer encoder with BERT pre-training parameters versus search domain-specific product category classification parameters.

\paragraph{\textbf{Query Evaluation}} In addition to checking validation accuracy to gauge overall performance of our models with only specific layers fine-tuned, we investigated what the models were actually learning. 
Queries with obvious or head phrases in them are identified properly in all models. An example is any query containing ``help with'' tends to be identified to be a query with help intent. Where the deep learning models showed promise is when there was not an obvious term in the query indicating an intent unless you had prior experience with the products that were being searched for. A search for a particular author is not understood to imply adult content unless it is previously known author only publishes erotic content.


\begin{figure}[t]
    \begin{subfigure}[t]{0.32\textwidth}
        \includegraphics[width=114pt]{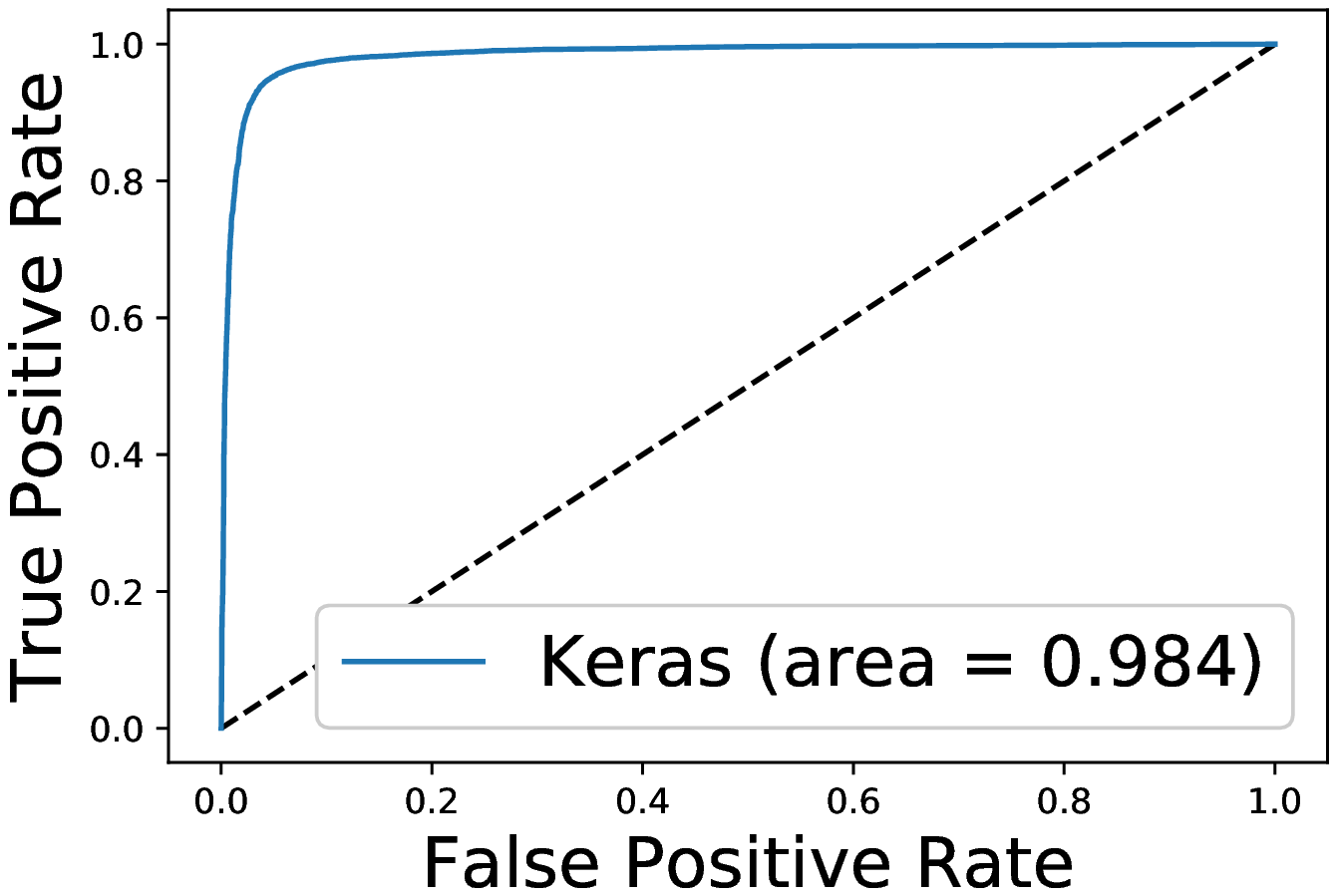}
        \caption{Adult}
    \end{subfigure}
    \begin{subfigure}[t]{0.32\textwidth}
        \includegraphics[width=114pt]{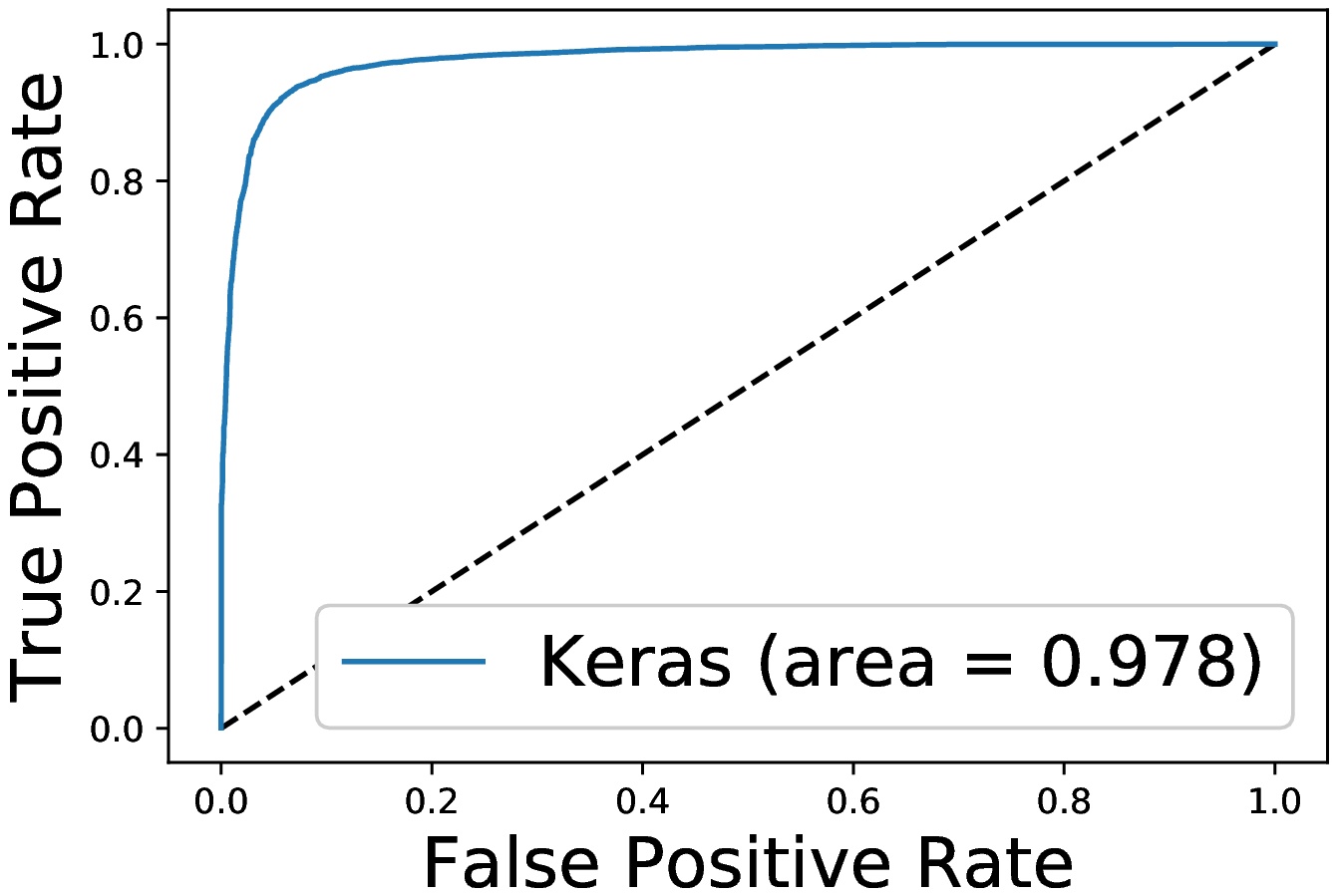}
        \caption{Help}
    \end{subfigure}
    \begin{subfigure}[t]{0.32\textwidth}
        \includegraphics[width=114pt]{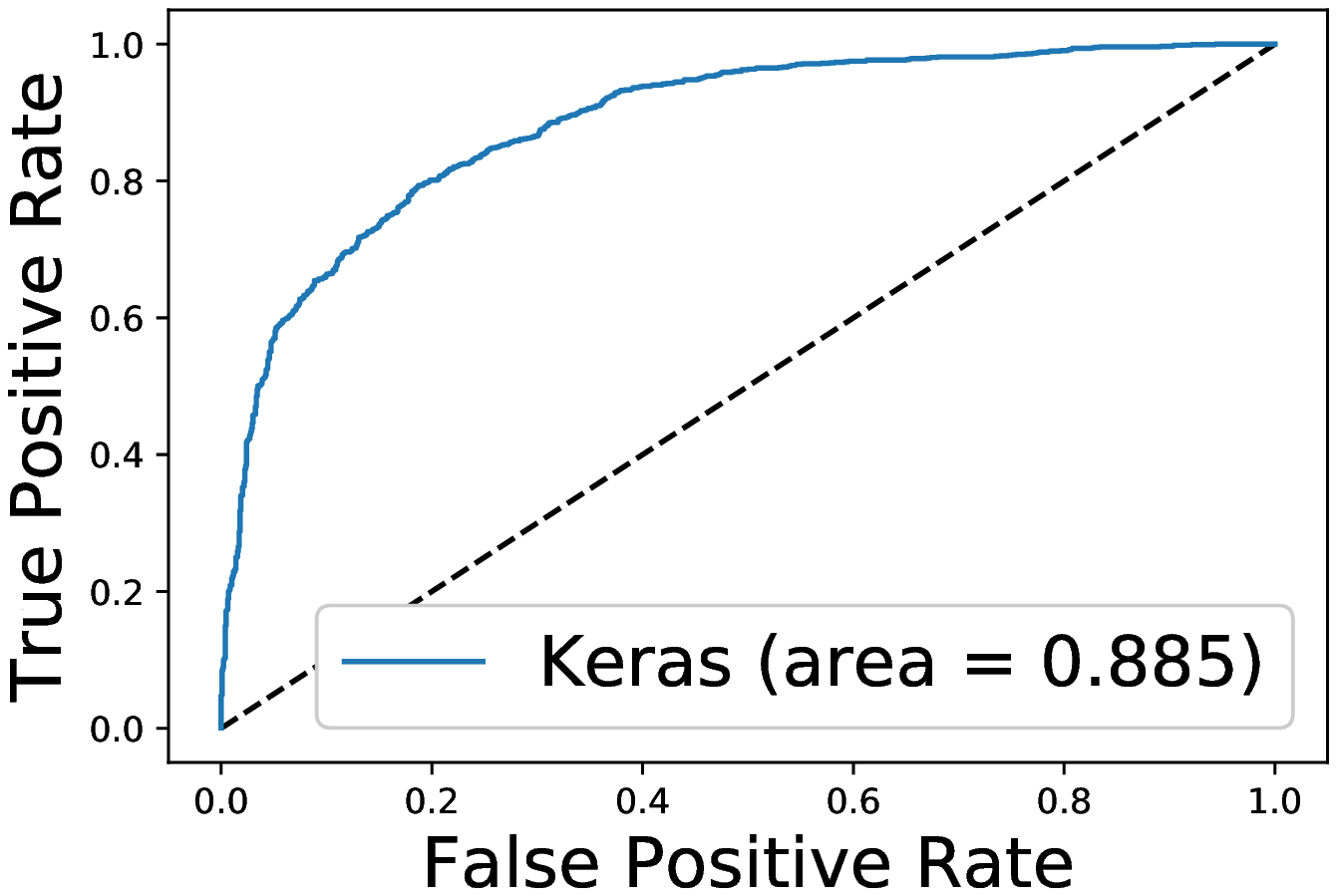}
        \caption{Low ASP}
    \end{subfigure}
    \caption{ROC Curves for Classifiers}
    \label{fig:roc}
\end{figure}

\paragraph{\textbf{Precision and Recall}} When calculating the label using a standard argmax objective, we found that the transformer encoder models sacrificed a bit of precision for big gains in recall when compared to approaches such as regular expression that are intended for high precision (such as using rules or regular expressions). \autoref{tab:results-test-set} indicates precision recall numbers when looking at the best performing models where at least the first 10 layers of the transformer encoder were frozen. The classification decision is made just by looking at the label with maximum value.



However, through basic tuning of classification value thresholds we are able to achieve even higher precision. For example, given the 2 element vector output from the classification layer for the adult classifier $[x,y]$, if we only regard a query as adult if the output for the adult label is 0.85 or more, we achieve a better precision (98.28\%) 
and with recall that is 87.49\%. 
\autoref{fig:roc} shows ROC curves. In practice, the classification threshold can be learnable to optimize for precision depending on the task or more sophisticated approaches can be used.

%% file: 06_production.tex
\section{Inference Considerations}
\label{sec:production}



When evaluating any model with more than 100M parameters, a major concern is inference latency. Tasks requiring near real-time performance would require a latency in the tens of milliseconds. We have seen a transformer encoder architecture as well as other large models perform great in an offline context but have only seen a handful of discussions, including bert-as-a-service \cite{bertservice} regarding the use of such models in a context with strict latency requirements. Basic latency tests looked promising with times ranging from 6 ms to 343 ms for tests with small (109MM parameters) or larger (330MM parameters) models when executed on EC2 instances ranging from p3.2xlarge to m5d.2xlarge. This demonstrates that achieving low latency for requests is possible with a model of this size when hosted on the latest GPU hardware.

The other major dimension of inference cost is GPU memory. The bigger a model, the more GPUs are required to host it and still meet throughput requirements. If we require 8 GPUs to handle a particular number of inference calls for a large model trained for a specific task, we would require double this number to host two tasks where each had a dedicated version of the model. At this rate, the cost of using large models in production could quickly become unacceptable. With parameter sharing we allow for the possibility that the same model on a constant number of GPUs can be used and re-used for an assortment of tasks without a rapid increase in infrastructure costs with the addition of new tasks.

\algblockdefx{FORALLP}{ENDFAP}[1]%
  {\textbf{for all }#1 \textbf{do in parallel}}%
  {\textbf{end for}}

\begin{algorithm} 
\caption{Inference Algorithm}
\label{alg:inference}



\textbf{procedure} $GetQueryIntents: (query)$\;

\begin{algorithmic} [H] 

\STATE $results \leftarrow []$
\STATE $(Name, Task, LayerSpec)[] \leftarrow QueryIntentSpecifications()$

\STATE $embeddings[] \leftarrow GetEmbeddings(query)$

\FORALLP{$name, task, layerspec \in (Name, Task, LayerSpec)$}
\STATE $results[name] \leftarrow task(query, embeddings[layerspec])$
\ENDFAP
\STATE \textbf{return} $results$

\end{algorithmic}
\end{algorithm}

\autoref{alg:inference} describes the basic call flow. \textit{GetEmbeddings} is a service call to the embedding service that is hosted on GPU boxes. The result passed back is a list of embeddings from final 3 layers of the transformer encoder. Remember during training that we found that freezing the first 10, 11, all layers, or none of the layers could produce the best results depending on the task (help, low ASP, adult). In order to make use of this at inference time we make a single call for a query to get back the embeddings for the various layers. We can then use this information accordingly in the task-specific calls described in \textit{GetQueryIntents} in \autoref{alg:inference}. Moreover, we cut down on latency by executing the retrieval of task-specific results in parallel. The list of query intents we care about is specified by the incoming request. We store service-side which encoder output layer needs to be used as input for each query intent task. This is represented by QueryIntentSpecifications in \autoref{alg:inference}. Each task corresponds to an inference call on a task specific model. We see in \autoref{tab:results-validation-set} that the number of task specific parameters required can be just a few thousand in some cases. In terms of required GPU inference hardware, the difference between serving 100 task specific transformer encoders and a single shared transformer encoder for a 100 tasks is enormous.

It is tempting to assume that more parameters tuned to a specific task is better. However, as we can see from \autoref{tab:results-validation-set}, more tuneable parameters does not necessarily mean better accuracy. Moreover, task-specific models, especially in the case where there are only a few thousand task-specific parameters in the model, can be hosted on CPU hardware which leads to additional infrastructure cost savings. We talked about basic approaches to classification in this paper (e.g. using a minimum threshold before accepting a classification decision) but more sophisticated approaches will also work.

We found that when leveraging an existing domain-specifically trained model we can build a production-grade classifier with far fewer training samples then was originally required during the more general domain-specific training step (\textless 3M samples vs 1-6B samples). This reduced data requirement combined with a fewer number of parameters needing to be trained allows new query intent classifiers to be built in a matter of hours on a single GPU. The cost of training and hosting additional downstream classification models is a fraction of what was required for the initial transformer encoder model.


%% file: 07_conclusion.tex
\section{Conclusions \& Future Work}
\label{sec:conclusion}

Recognizing and classifying search queries into granular classes of query intents helps a shopping search engine to customize the shopping experience delivered for each customer scenario. We discussed three examples of query intents in this paper but there are thousands of other intents a user might have. Knowing these intents can allow the retrieval of more relevant products based on category, customization of ranking functions to show the most relevant products first, and customized user interfaces based on intents.

Building a framework that supports thousands of shopping intent classifiers in a cost-effective way is both challenging and expensive. Sharing various types of in-domain data, computational resources, and models in production when possible is a must. We have demonstrated that transformer encoders can have shareable components that can be leveraged by a number of downstream classifiers. Hosting the shareable components as a service so that a common layer output can be used as input for other smaller task-specific models allows us to quickly train and deploy classifiers with a minimal number of additional parameters. We obtain the benefits of applying a large model to search query understanding tasks but mitigate the infrastructure cost and requirements of hosting a dedicated large model for each individual task.